\documentclass[aps,prd,showpacs,preprintnumbers,groupedaddress]{revtex4}
\usepackage{amsmath}
\usepackage{graphicx}
\usepackage{bm}
\begin{document}
\preprint{YITP-SB-02-47}
\preprint{BNL-HET-02/19}
\preprint{hep-ph/0208262}
\title{Distinguishing scalar from pseudoscalar Higgs production at the LHC}
\author{B. Field}
\email[]{bfield@ic.sunysb.edu}
\affiliation{C.N. Yang Institute for Theoretical Physics, Stony Brook University, 
             Stony Brook, New York 11794-3840, USA}
\affiliation{Department of Physics, Brookhaven National Laboratory,
             Upton, New York 11973, USA}
\date{August 29, 2002}
\begin{abstract}
In this letter we examine the production channels for the scalar or pseudoscalar Higgs 
plus two jets at the CERN Large Hadron Collider (LHC). We identify possible signals for 
distinguishing between a scalar and a pseudoscalar Higgs boson.
\end{abstract}
\maketitle

\section{Introduction}
The Higgs mechanism is responsible for electroweak symmetry breaking in the Standard 
Model (SM). The experimental lower limit on the Higgs mass is approximately 
114~GeV\cite{lep}. There are many models that contain more than one Higgs boson in 
various numbers of doublets. In the Minimal Supersymmetric Standard Model (MSSM) there 
are two Higgs doublets that give five physical Higgs bosons: two neutral $(H^0_1, 
H^0_2)$, two charged $H^\pm$, and one neutral pseudoscalar $A$ (for review see 
\cite{ghkd}). In the MSSM the mass limits change slightly with the lightest of the two 
neutral scalars $H^0_1$ (afterwards reffered to as simply H) having a mass greater that 
about $91$~GeV and the pseudoscalar being more massive than roughly $92$~GeV\cite{mssm}.

Finding one or more Higgs bosons is the top priority of high energy physics programs 
around the world. A subset of Higgs bosons in some doublet models may be experimentally 
difficult to distinguish. The characteristics of the scalar $H$ and pseudoscalar $A$ Higgs 
boson within the MSSM are of particular interest.

We study the production of both a scalar and pseudoscalar Higgs in association with two 
jets in hadron collisions. At the LHC the primary processes that produce a Higgs plus 
two jets are $gg \rightarrow ggH$ and $qg \rightarrow qgH$, accounting for approximately 
60\%(40\%) of the total cross-section respectively. The same is true for the production 
of the pseudoscalar. Other channels that contribute to the total cross-section include 
$qq \rightarrow ggH$ and $qq \rightarrow qqH$, although these channels have been shown to 
add very little to the total cross-section. In the following calculations, only the two 
dominant channels were considered as the other channels are negligible.

Total cross-sections of the scalar and pseudoscalar plus two 
jets exist\cite{russel,russel2,higgscross} at the lowest order.
Total cross-sections for the inclusive production have been
calculated at NLO for the scalar \cite{sally,spira1} and for the pseudoscalar
\cite{schaffer} and at NNLO for the scalar
\cite{harkil2,harkil3,catani1,catani2,anast2} and for the pseudoscalar
\cite{harkil1,anast1}. If we define the K-factor to be the ratio of the higher order
cross-section to the lowest order, the rate increase at the LHC at NNLO
for the scalar inclusive 
processes\cite{harkil2} was reported to be $\mathrm{K}^\mathrm{NNLO}(\mathrm{pp}\rightarrow
\mathrm{H+X}) = 2-2.2$ and for pseudoscalar the K-factor\cite{harkil3} can be determined to be 
$\mathrm{K}^\mathrm{NNLO}(\mathrm{pp}\rightarrow \mathrm{A+X}) = 2-2.3$
in the mass range $M_{H,A}=100 - 200$~GeV. The total cross-section and the differential  
cross-section for a scalar Higgs plus one jet has been calculated by
\cite{catani1,catani2,jack,higgscross,florian} and the total rate was also shown to increase 
substantially. The NLO corrections to pseudoscalar plus one jet have not yet been 
computed. In all of the processes cited above the rates increased by comparable amounts. We 
expect our estimates of the Higgs plus two jets  rates to be 
conservative, however, since our proposed observable is normalized to the cross-section, we 
do not expect major changes to occur in our analysis at higher orders.

In this letter, we propose a technique for distinguishing between a scalar and a 
pseudoscalar Higgs when produced in association with two jets by means of a splitting 
that occurs in a specific integrated operator moment. This distinction is important both 
experimentally and theoretically in order to separate the two kinds of events and 
understand the properties of these particles which would otherwise be very difficult due 
to the similarity in their physical observables.

\section{Effective Lagrangian}
We work in the limit that the top quark is much heavier that the Higgs 
boson\cite{vain,sally,spira1,spira2,schaffer,spira3}, 
integrating out the top quark and neglecting all the other quarks that would normally 
appear in the loop diagrams. This has been shown to be an excellent approximation and 
remains very good even when the Higgs mass is heavier that the mass of the top quark. In 
general this approximation is considered to be a good one when $M_{H,A} < 2m_t$. We 
consider Higgs bosons lighter than $200$~GeV. The effective Lagrangian used in the 
scalar case is defined as
\begin{equation}
{\cal L}^H_{\mathrm{eff}} = - \frac{1}{4}g_H H G^a_{\mu \nu} G^{a, \mu \nu}
\end{equation}
where $g_H = \alpha_s/3 \pi v$. $G^a_{\mu \nu}$ is the field-strength tensor for the 
gluons. The vacuum expectation value (vev) of the Higgs field is determined in the usual 
way as $v^2 = (\sqrt{2}G_F)^{-1}$ and is numerically equal to approximately $246$~GeV. 
For the pseudoscalar case we let the Higgs couple to the quarks with a $\gamma_5$ and the 
effective Lagrangian\cite{footnote}can be written as
\begin{equation}
{\cal L}^A_{\mathrm{eff}} = \frac{1}{4}g_A A G^a_{\mu \nu} \tilde{G}^{a, \mu \nu}
\end{equation}
where $g_A = \alpha_s/2 \pi v$. Here 
$\tilde{G}^a_{\mu \nu} = 1/2 \epsilon^{\mu \nu \rho \sigma}G^a_{\rho \sigma}$ is the dual 
of the gluon field-strength tensor.

This effective Lagrangian generates a scalar Higgs coupling to two, three, and four 
gluons or a pseudoscalar coupling to two or three gluons. The four gluon coupling to a 
pseudoscalar vertex vanishes via the Jacobi identity as it is proportional to a 
completely antisymmetric combination of structure constants. The Feynman rules for these 
effective theories can be found in \cite{russel} (for the scalar) and \cite{russel2} (for 
the pseudoscalar).

\section{Observables and Moments}
We present our results for the LHC with $\sqrt{S}=14$~TeV. We have used the CTEQ6L parton 
distribution functions\cite{cteq} with $\Lambda_5^\mathrm{LO}=226$~MeV with a one-loop 
running of $\alpha_s$ for consistency with a value of $\alpha_s(M_Z)=0.137$. The 
transverse momentum ($p_\mathrm{t}$) was constrained to be more than $25$~GeV for 
the Higgs and each of the two jets. Also the rapidity was constrained to be $|y| < 
2.5$ for all the outgoing particles. The separation of the jets was restricted to be
$\Delta R \equiv \sqrt{\Delta \phi^2 + \Delta \eta^2} \ge 0.7$.

\begin{figure}
\includegraphics{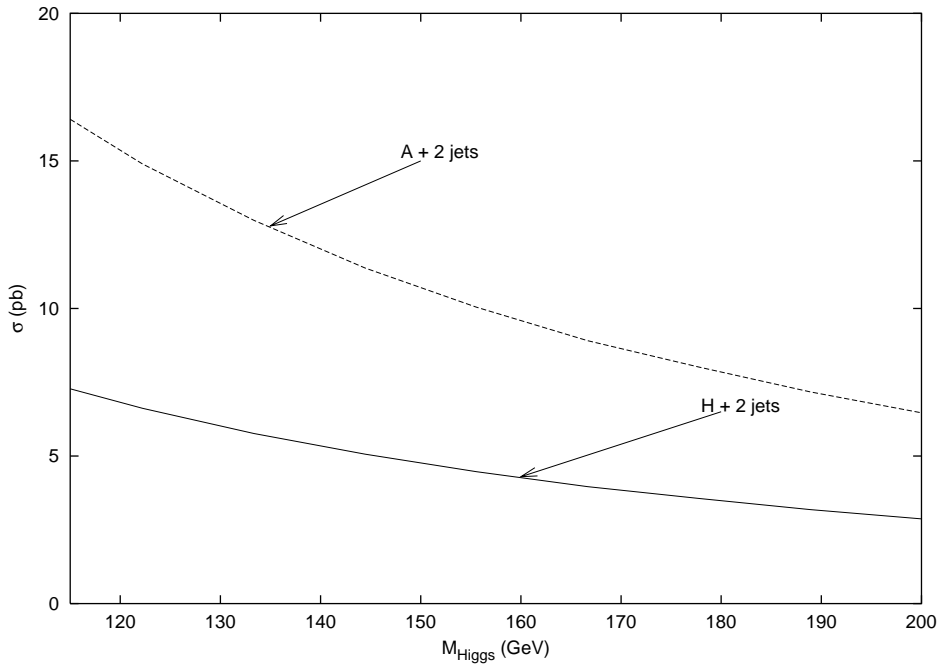}%
\caption{Total cross-sections for scalar and pseudoscalar Higgs plus two jets. These 
curves are for the LHC with the cuts described in the text.\label{cross}}
\end{figure}

The total cross-section of these two channels are shown in Fig.~\ref{cross}. These 
cross-sections agree exactly with those in the literature\cite{russel,russel2} once 
the problems with the effective coupling constants are remedied. When 
plotted in this linear fashion it is interesting to note the differences in the 
dependence of the cross-sections on the mass of the Higgs boson. Both total 
cross-sections loose more than two-thirds of their value from $100 - 200$~GeV and appear 
in the approximate ratio of $(g_H/g_A)^2 = 4/9$ due to the similarity in their 
matrix elements.

\begin{figure}
\includegraphics{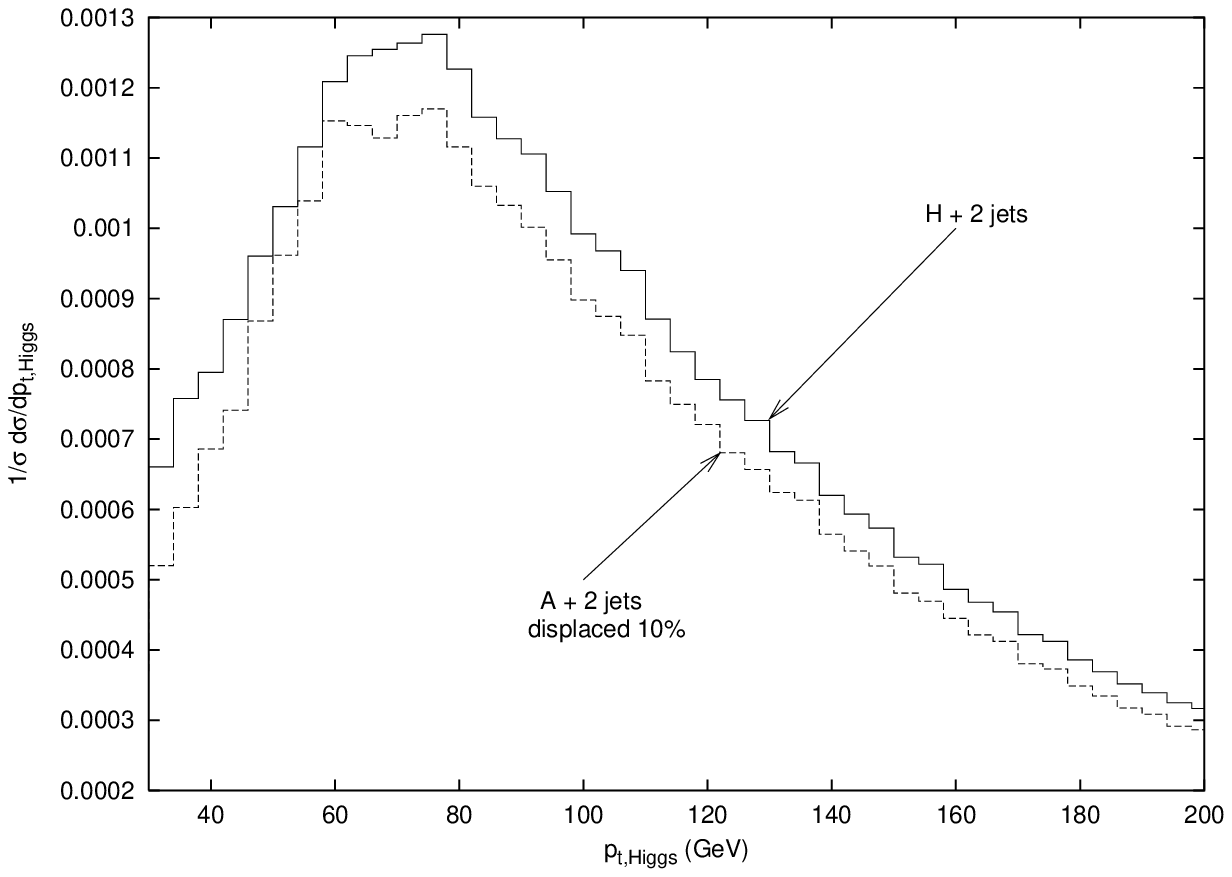}%
\caption{Normalized transverse momentum spectrum of the scalar or pseudoscalar Higgs 
production channels plus two jets. The Higgs mass for both the scalar and pseudoscalar is 
$120$~GeV. Note that the pseudoscalar Higgs has been displaced down by $10$\% to allow 
the two curved to be distinguished. These curves are for the LHC with the cuts described 
in the text.\label{pt_spectrum}}
\end{figure}

\begin{figure}
\includegraphics{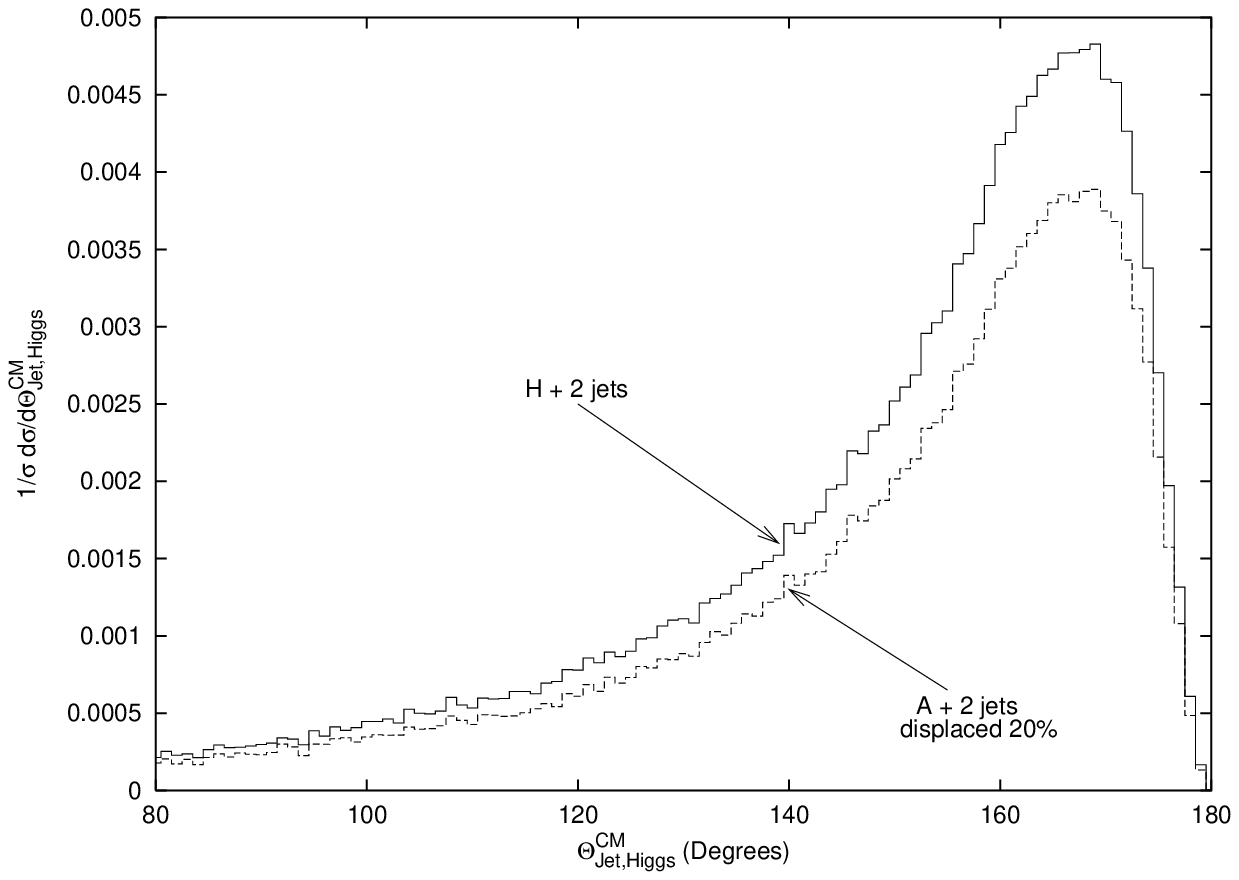}%
\caption{Normalized opening angle in the center-of-momentum frame between the Higgs and 
the highest $p_\mathrm{t}$ jet for the scalar and pseudoscalar Higgs production 
channels plus two jets. The Higgs mass for both the scalar and pseudoscalar is 
$120$~GeV. Note that the pseudoscalar Higgs has been displaced down by $20$\% to allow 
the two curved to be distinguished. These curves are for the LHC with the cuts described 
in the text.\label{angle}}
\end{figure}

Fig.~\ref{pt_spectrum} shows the normalized transverse momentum spectrum of both the 
production channels. The pseudoscalar Higgs $p_\mathrm{t}$ spectrum was displaced down by 
$10$\% to allow the two curves to be distinguished. If this had not been done, the 
curves would lie virtually on top of one another. Fig.~\ref{angle} shows the 
center-of-momentum angle between the Higgs and the highest $p_\mathrm{t}$ jet for 
the two reactions. This shows what would be expected na\"{\i}vely, that the Higgs 
prefers to come out back-to-back with the highest $p_\mathrm{t}$ jet. Once again, 
the pseudoscalar curve has been scaled down by $20$\% to allow both curves to be 
seen clearly. No significant differences between these curves were found. 

The authors of \cite{gunion} presented a technique for determining the $CP$ nature 
of the Higgs boson in $t\bar{t}H$ production based on certain weighted moments of 
the cross-section. The cross-section integral was weighted by operators ${\cal 
O}_{CP}$. The six operators presented in \cite{gunion} are scalar and cross products 
of the momentum of the outgoing particles (in this case the massive top quarks). We 
propose using the same test for the massless quarks and gluons that make up the 
jets. All of these weighted moments were examined as well as some novel ones and the 
only operator from these sets that produced a significant difference between the 
scalar and the pseudoscalar signals was the operator\cite{gunion}
\begin{equation}
a_1 = \frac{ (\vec{p_1} \times \widehat{z}) \cdot (\vec{p_2} \times \widehat{z}) }
           {|(\vec{p_1} \times \widehat{z}) \cdot (\vec{p_2} \times \widehat{z})|}
\end{equation}
when it was integrated and normalized as prescribed below
\begin{equation}
\alpha[{\cal O}_{CP}] \equiv \frac{1}{\sigma} \int {\cal O}_{CP} \, d\sigma \, dPS
\end{equation}
where $p_1$ and $p_2$ are the momentum of the two jets and $\widehat{z}$ is the axis 
of the beam. The $a_1$ operator is sensitive to the cosine of the angle between the 
transverse momentum vectors of the two jets. Distinguishing between the two jets is 
not important as this moment is invariant under $1 \leftrightarrow 2$. Another 
combination of momentum in the above equations that was considered was to use the 
moment operators presented in \cite{gunion} with $p_1 = p_\mathrm{Higgs}$ and $p_2$ 
the momentum of the highest $p_\mathrm{t}$ jet. However, this yielded no differences 
in the integrated moments making this definition of little use for these channels.

\begin{figure}
\includegraphics{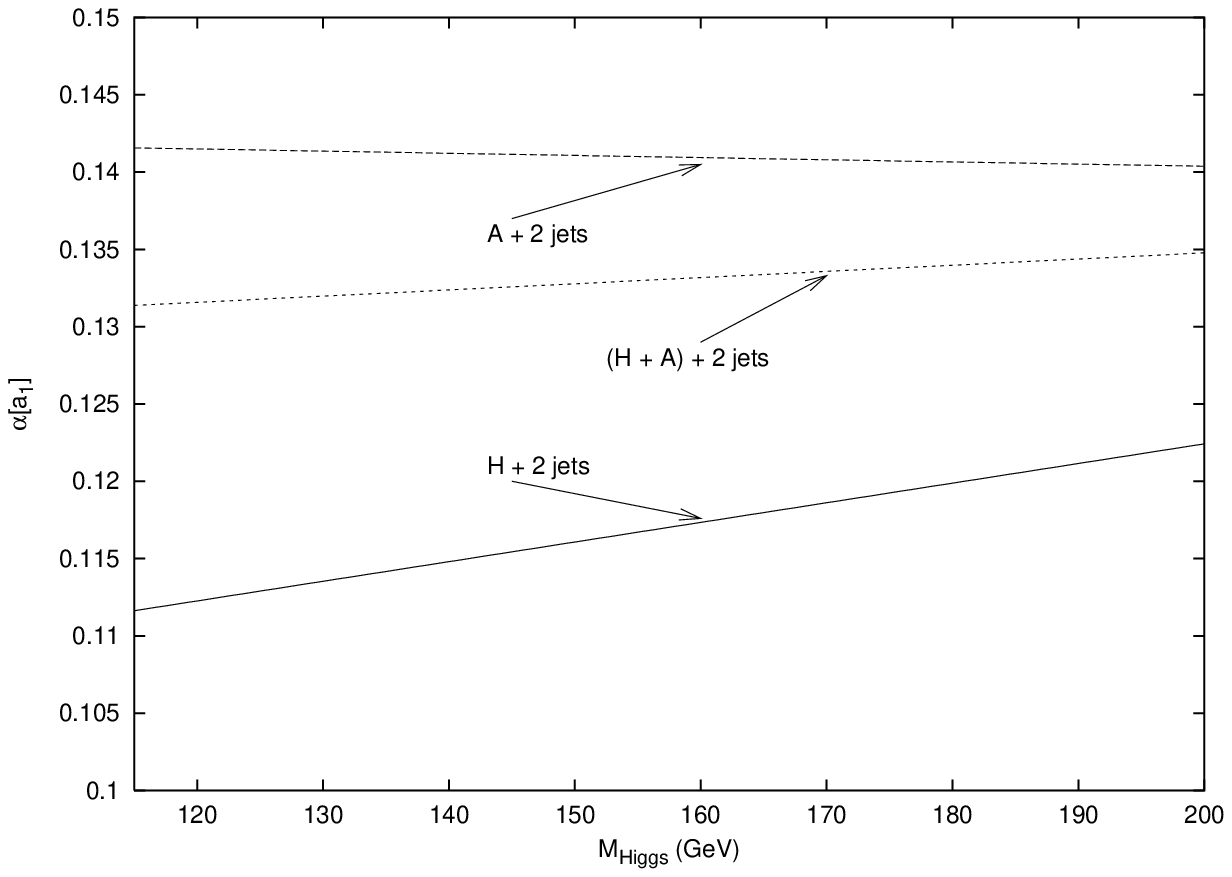}%
\caption{Normalized integrated moment $\alpha [a_{1}]$ for the scalar and pseudoscalar 
Higgs plus two jets. The center curve shows what the observable would look like if both
of the processes were to be measured at the same time with degenerate masses. The splitting 
between the two different production channels is clear at all mass scales with a 
statistical uncertainty of about $5$\%. These curves are for the LHC with the cuts described in the 
text.\label{cp_a1}}
\end{figure}

Fig.~\ref{cp_a1} shows the results of this integration as a function of the Higgs 
mass. If we consider a conservative estimate of $100$ fb$^{-1}$ of integrated 
luminosity at the LHC and take a branching ratio of approximately $10^{-3}$ 
as an order of magnitude for the decay of the Higgs to a pair of photons, then 
the $\alpha [a_{1}]$ observable will have a statistical uncertainty of about 
$5$\%, making these two signals distinguishable at all mass scales. With this 
conservative estimate on the integrated luminosity we would expect to see about 
$600$ scalar events and $1000$ pseudoscalar events for a Higgs mass of $120$~GeV
in this channel for this Higgs decay. These numbers are supported by a more 
detailed analysis using the actual branching ratios calculated using 
\textsc{hdecay}\cite{hdecay} 
in the $\tan\beta=1$ limit for the pseudoscalar.

This integrated moment showed a modest ($30$\%) splitting at all Higgs mass scales from 
$100 - 200$~GeV. The pseudoscalar does not show much mass dependence. However, the 
scalar integrated moment rises slightly with increasing Higgs mass. This effect 
might also be useful as another method for constraining the mass of the scalar 
Higgs boson. The splitting in Fig.~\ref{cp_a1} helps to remove the problems created 
by the degeneracy in the physical observable of the scalar and the pseudoscalar. If 
the two signals could not be separated, the doublet structure of the model would not 
be easily measured. In the case of the MSSM this would mean that part of the 
supersymmetric signal might be lost or the mass of the scalar Higgs may be 
determined incorrectly if the pseudoscalar events were wrongly identified as scalar 
events.

Separating the two signals is theoretically intriguing because it appears to be one 
of the only ways to predict a difference between the scalar and pseudoscalar 
events of this nature at the LHC by means other than the magnitude of their 
cross-section. This is also interesting experimentally as it leads to the 
possibility of separating the two kinds of Higgs events with the added bonus that 
the $z$ momentum is not needed in this analysis.

\section{Conclusions}
The production channels of the scalar or pseudoscalar Higgs plus two jets were found 
to have many similarities in their physical observables and one important difference 
in the integrated moment $\alpha[a_1]$. This may help to reduce the difficulty in 
distinguishing between the two types of events at the LHC. The most important aspect 
of separating the two signals is to make sure that the doublet structure (the 
supersymmetric signal in the case of the MSSM) is not lost because of its small 
cross-section and its similarity to the scalar Higgs with respect to its physical 
observables or wrongly determining the mass of the scalar Higgs by misidentifying 
pseudoscalar events as scalar events. The proposed technique presented in this 
letter may enable these two signals to be separated after a full detector simulation 
is preformed.

\appendix
\section{Differences in the Amplitudes}
It turns out that the differences in the scalar (or pseudoscalar) plus two jets 
amplitudes squared were very small. The differences will be presented using the 
helicity basis presented in \cite{russel, russel2} to make for the most compact 
matrix elements squared. These matrix elements have been found to be in exact 
analytic agreement with the four dimensional matrix elements presented in 
\cite{jack}. We identify the momentum as follows (where $X$ should be considered the 
Higgs for the process in question, playing the part of either the scalar or the 
pseudoscalar). All the momenta are outgoing.
\begin{align}
q(p_1) + \bar{q}(p_2) &\rightarrow g(-p_3) +       g(-p_4) + X(-p_5) \\
g(p_1) +       g(p_2) &\rightarrow g(-p_3) +       g(-p_4) + X(-p_5) \\
q(p_1) + \bar{q}(p_2) &\rightarrow q(-p_3) + \bar{q}(-p_4) + X(-p_5).
\end{align}

In the following we define 
$S_\mathrm{ab} = (p_\mathrm{a} + p_\mathrm{b})^2 = 2p_\mathrm{a}\cdot p_\mathrm{b}$. 
Color factors have been included in the expression for the $qqggH(A)$ and $qqqqH(A)$ 
channels as they affect the terms differently but not in the expression for the 
$ggggH(A)$ channel as there is one overall color factor for all the matrix elements 
squared. Here $N$ is the number of colors. Color and spin averages have not been 
included nor have any coupling constants.

For the $qqggH(A)$ channel the difference in the scalar minus the pseudoscalar 
amplitude squared was $15$ terms out of $626$. Setting the color factors to match 
those presented in \cite{jack}, $C_O = (N^2-1)/N$ and $C_K = (N^2-1)N$ the 
difference was found to be
\begin{align} \nonumber
|{\cal M}|^2_{qq \rightarrow ggH} - |{\cal M}|^2_{qq \rightarrow ggA} &= 
2C_K - 6C_O \\ \nonumber
+ \biggr( \biggl\{ \frac{4C_O}{S_{12}^2S_{34}^2} &\biggl[ S_{13}S_{14}S_{23}S_{24} - 
S_{13}^2S_{24}^2 \biggl] + 4C_O\frac{S_{13}S_{24}}{S_{12}S_{34}} \\
\frac{1}{S_{13}S_{24}} &\biggl[ C_O(S_{12}S_{34} - S_{14}S_{23}) + C_K(S_{14}S_{23} 
- S_{12}S_{34}) \biggr] \biggr\} + \{ 3 \leftrightarrow 4 \} \biggr).
\end{align}

For the $ggggH(A)$ channel the difference in the scalar minus the pseudoscalar was 
$16$ terms out of $2761$. The overall color factor is $N^2(N^2-1)$. The difference 
was
\begin{align} \nonumber
|{\cal M}|^2_{gg \rightarrow ggH} - |{\cal M}|^2_{gg \rightarrow ggA} = 48 +
\biggl( 8 \biggl\{ 
  \frac{1}{2} &\frac{1}{S_{12}^2S_{34}^2}\biggl[S_{13}S_{24} - S_{14}S_{23}\biggr]^2 
- \frac{1}{2}  \frac{1}{S_{12}  S_{34}  }\biggl[S_{13}S_{24} + S_{14}S_{23}\biggr]^2 \\
+&\frac{1}{S_{13}^2S_{24}^2} \biggl[ S_{12}S_{34} - S_{14}S_{23} \biggr]^2 -
  \frac{1}{S_{13}  S_{24}  } \biggl[ S_{12}S_{34} + S_{14}S_{23} \biggr]
\biggr\} + \{ 3 \leftrightarrow 4 \} \biggr).
\end{align}

Finally, there are two cases for the $qqqqH(A)$ amplitude squared. If there are 
identical quarks allowed in the scattering process ($q\bar{q}q\bar{q}H(A))$ then 
there are two diagrams that contribute. The color factors here are $C_A=N$ and $C_F 
= (N^2-1)/2N$. The difference in the amplitudes squared is $19$ out of $39$ terms 
and is equal to
\begin{align}
|{\cal M}|^2_{qq \rightarrow qqH} - |{\cal M}|^2_{qq \rightarrow qqA} &=
4C_AC_F \biggl(2 - \frac{4}{C_A} +  \frac{(S_{13}S_{24} - S_{14}S_{23})^2}
{S_{12}^2S_{34}^2} + \frac{(S_{14}S_{23}-S_{12}S_{34})^2}{S_{13}^2S_{24}^2} \\
&- 2\frac{S_{13}S_{24}}{S_{12}S_{34}} - 2\frac{S_{12}S_{34}}{S_{13}S_{24}}
+\frac{2}{C_A}\biggl( \frac{S_{12}S_{32}-S_{14}S_{23}}{S_{13}S_{24}}
\biggl) + \{ 3 \leftrightarrow 4 \} \biggr).
\end{align}

If a different quark pair is created ($q\bar{q}q'\bar{q}'H(A))$, then the difference 
is smaller as only one diagram is needed for the amplitude. Here $6$ out of $10$ 
terms survive and are equal to
\begin{equation}
|{\cal M}|^2_{qq \rightarrow qqH} - |{\cal M}|^2_{qq \rightarrow qqA} =
4C_F\biggl(1 + \biggl\{ \frac{(S_{13}S_{24}-S_{13}S_{24})^2}{S_{12}^2S_{34}^2}
- \frac{S_{13}S_{24} + S_{14}S_{23}}{S_{12}S_{34}} \biggr\}
+ \{ 3 \leftrightarrow 4 \} \biggr).
\end{equation}
It should also be noted that all these differences are invariant under $1 
\leftrightarrow 2$. 

\begin{acknowledgments}
The author would like to thank J.~Smith, S.~Dawson, R.P.~Kauffman, S.V.~Desai, 
W.~Kilgore, F.~Paige, and J.~Laiho for their help and comments on this paper at all 
stages of its development. The author was partially supported by the National Science 
Foundation grant PHY-0098527 and the U.S. Department of Energy under Contract No. 
DE-AC02-98CH10886.
\end{acknowledgments}


\begin{thebibliography}{2002}
\bibitem{footnote}
There is some confusion over the coupling constant
for the pseudoscalar case in the literature. The correct coupling is found in
Ref.~\cite{schaffer}. There is an extra factor of $1/4$ in Ref.~\cite{russel2} leading
to a cross-section $16$ times too small for the pseudoscalar case. It seems that the
$1/4$ from the effective Lagrangian was incorporated into the coupling constant
by mistake. The Feynman rules in both papers are correct if the coupling constant
from Ref.~\cite{schaffer} paper is used.
\end{thebibliography}
\end{document}